\documentclass[iop,numberedappendix,twocolappendix,appendixfloats]{emulateapj}
\usepackage{graphicx,subfigure,amsmath, amsfonts, amssymb,aas_macros,times}
\usepackage[usenames,dvipsnames]{color}
\usepackage[export]{adjustbox}
\usepackage[bookmarks=false]{hyperref}
\hypersetup{
  colorlinks= true, 
  urlcolor  = MidnightBlue, 
  filecolor=black,
  runcolor=blue,
  linkcolor = Maroon, 
  citecolor = BlueViolet 
} 

\newcommand{\atlas}{ATLAS$^{\rm 3D}$}

\newcommand{\kms}{km~s$^{-1}$}

\newcommand{\twco}{\hbox{$^{12}$CO}}
\newcommand{\thco}{\hbox{$^{13}$CO}}

\newcommand{\rco}{$\mathcal{R_{\rm 12/13}}$}

\shorttitle{Star Formation and Molecular Gas in NGC\,5195}
\shortauthors{Alatalo et al.}

\begin{document}

\title{After the interaction: an efficiently star-forming molecular disk in NGC\,5195}
\author{Katherine Alatalo,$^{1}$\altaffilmark{$\dagger$} Rebeca Aladro,$^{2,3}$ Kristina Nyland,$^{4}$ Susanne Aalto,$^{3}$ Theodoros Bitsakis,$^{5}$ John S. Gallagher$^{6}$ \& Lauranne Lanz$^{7}$}

\affil{
$^{1}$Observatories of the Carnegie Institution of Washington, 813 Santa Barbara Street, Pasadena, CA 91101, USA\\
$^{2}$European Southern Observatory, Avda. Alonso de C\'o—rdova 3107, Vitacura, Santiago, Chile\\
$^{3}$Department of Earth and Space Sciences, Chalmers University of Technology, Onsala Observatory, 439 94 Onsala, Sweden\\
$^{4}$National Radio Astronomy Observatory, 520 Edgemont Road, Charlottesville, VA 22903, USA\\
$^{5}$Instituto de Radioastronom\'ia y Astrof\'isica, Universidad Nacional Aut\'onoma de M\'exico, C.P. 58190, Morelia, Mexico\\
$^{6}$Department of Astronomy, University of Wisconsin-Madison, 5534 Sterling, 475 North Charter Street, Madison WI 53706, USA\\
$^{7}$Infrared Processing and Analysis Center, California Institute of Technology, MC100-22, Pasadena, CA 91125, USA
}
\altaffiltext{$\dagger$}{Hubble fellow}
\email{kalatalo@carnegiescience.edu}
\slugcomment{Accepted by the Astrophysical Journal, Aug 3, 2016}





\begin{abstract}
We present new molecular gas maps of NGC\,5195 (alternatively known as M51b) from the Combined Array for Research in Millimeter Astronomy (CARMA), including \twco(1--0), \thco(1--0), CN(1$_{0,2}$--$0_{0,1}$), CS(2--1), and 3mm continuum. We also detected HCN(1--0) and HCO$^+$(1--0) using the Onsala Space Observatory. NGC\,5195 has a \twco/\thco\ ratio (\rco\,=11.4\,$\pm$\,0.5) consistent with normal star-forming galaxies. The CN(1--0) intensity is higher than is seen in an average star-forming galaxy, possibly enhanced in the diffuse gas in photo-dissociation regions.  Stellar template fitting of the nuclear spectrum of NGC\,5195 shows two stellar populations: an 80\% mass fraction of old ($\gtrsim$\,10\,Gyr) and a 20\% mass fraction of intermediate-aged ($\approx$\,1\,Gyr) stellar populations, providing a constraint on the timescale over which NGC\,5195 experienced enhanced star formation during its interaction with M51a.  The average molecular gas depletion timescale in NGC\,5195 is $\langle\tau_{\rm dep}\rangle$\,=\,3.08 Gyr, a factor of $\approx$2 larger than the depletion timescales in nearby star-forming galaxies, but consistent with the depletion seen in CO-detected early-type galaxies. While radio continuum emission at centimeter and millimeter wavelengths is present in the vicinity of the nucleus of NGC\,5195, we find it is most likely associated with nuclear star formation rather than radio-loud AGN activity. Thus, despite having a substantial interaction with M51a $\sim$1/2\,Gyr ago, the molecular gas in NGC\,5195 has resettled and is forming stars at an efficiency consistent with settled early-type galaxies at the present time.
\end{abstract}

\keywords{
galaxies: elliptical and lenticular, cD -- galaxies: individual: NGC\,5195 -- galaxies: interactions -- galaxies: ISM -- radio lines: galaxies}

\section{Introduction}
\label{sec:intro}
Galaxy populations in the present-day universe largely populate a bimodal distribution between blue and red optical colors with few objects occupying an intermediate color "green valley" \citep{baade58,holmberg58,tinsley78,strateva+01,baldry+04}. Slowly evolving galaxies would not necessarily pass rapidly through a phase with intermediate colors. Thus the slow overall evolution of galaxies since {\em z}\,$\sim$\,1 in combination with the increasing fraction of galaxies on the "red sequence" suggests that galaxies in the green valley experienced a rapid truncation or quenching of their star formation \citep{bell+03,faber+07}.  However, the existence of spirals with intermediate colors that appear to be evolving at secular rates suggests the existence of an alternate path to the green valley. The intermediate colors in these spirals could result from the steady build-up long-lived lower mass stars leading to a more leisurely transition through the green valley. Two types of galaxies with intermediate colors therefore may exist, those produced by slow evolution and rapid quenching events, that cannot be separated on the basis of broad-band optical colors alone.

Other types of observations, however, can be used to separate these two types of green valley galaxies. Systems where star formation rates (SFRs) have rapidly changed show distinct signposts, including unusual infrared colors. \citet{johnson+07} and \citet{walker+10} were able to show that the {\em Spitzer} colors of Hickson Compact Group galaxies (a known rapidly transitioning population; \citealt{hickson82}) showed a dearth of intermediate {\em Spitzer} IRAC colors \citep{lacy+04}, suggestive of rapid transformation. While the galaxy population as a whole did not show bimodality in {\em Spitzer} colors, it did strongly bifurcate using {\em WISE} [4.6]--[12]$\mu$m colors \citep{ko+13,yesuf+14,a14_irtz}.  In the modern universe ({\em z}\,$\sim$\,0), this transformation appears to be permanent \citep{appleton+14,young+14}, thus understanding all pathways that can lead a blue late-type galaxy to become a red early-type is essential.

\begin{figure*}
\includegraphics[width=0.99\textwidth]{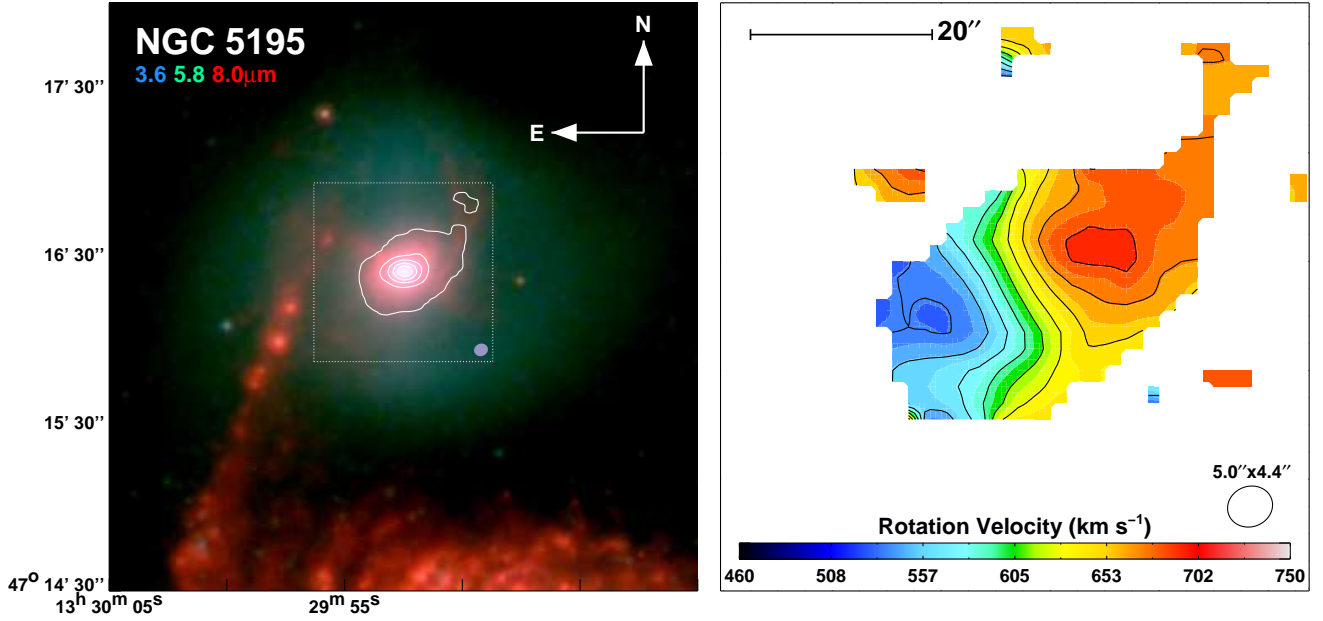} \vskip -3.5mm
\caption{{\bf(Left):} The 3-color 3.6$\mu$m\,5.8$\mu$m\,8.0$\mu$m {\em Spitzer} IRAC image of NGC\,5195 overlaid with the moment0 map for \twco(1--0) (white contours).  The CARMA beam (light purple) is shown at the bottom right of the enclosed box. The red emission represents nonstellar 8.0$\mu$m emission, from the 7.7$\mu$m PAH feature, with a spiral arm of M51a shown in the foreground. The most prominent PAH emission associated with NGC\,5195 is traced by the \twco, including the filament extending to the upper right. {\bf(Right):} A zoomed-in box of NGC\,5195 shows the \twco\ moment1 map. All moment1 and channel maps follow the same color scale. The extended filament is also visible in the moment1 map, and on average NGC\,5195 appears to follow regular rotation.}
\label{fig:pah+co}
\end{figure*}

Until recently, it was assumed that in order for a galaxy to metamorphose, it would have to shed its interstellar medium as it quenched star formation \citep{hopkins+06}, but evidence is mounting that this is not strictly necessary. Through the \atlas\ survey, \citet{young+11} showed that at least 22\% of early-type galaxies contain a molecular reservoir, though for the most part the molecular gas fraction of these systems did not exceed 1\% (save for the extraordinary galaxy NGC\,1266; \citealt{alatalo+11}).  \citet{french+15} and \citet{rowlands+15} showed that there were also substantial molecular reservoirs in poststarburst galaxies, a subset of objects known to have rapidly quenched their star formation \citep{dressler+gunn83,zabludoff+96,quintero+04}. Shocked poststarburst galaxies \citep{a16_sample}, identified based on evidence of intermediate age stars and a lack of star formation from their ionized gas ratios also contain substantial gas reservoirs \citep{a16_spogco}, despite being suspected to have started their metamorphosis from blue, star-forming spiral to red, quiescent early-type galaxies.

 \citet{a15_hcgco} surveyed the molecular gas content in transitioning Hickson Compact Group galaxies \citep{cluver+13,lisenfeld+14} using the Combined Array for Research in Millimeter Astronomy (CARMA; \citealt{carma})\footnote{http://www.mmarray.org} and found that the molecular gas fractions in this subset of transitioning galaxies are not related to their color (and thus, the transition phase), but rather the star formation efficiency is related to the color. \citet{a15_hcgco} suggest that the inability for the molecular gas to form stars efficiently has more influence on a galaxy's path onto the red sequence than the total content of molecular material available. Studies by the \atlas\ team seem to support this, with early-type galaxies showing slightly suppressed star formation as well \citep{davis+14}, although a direct color comparison has yet to be done on these sources.
\begin{figure*}[t]
\includegraphics[width=0.99\textwidth]{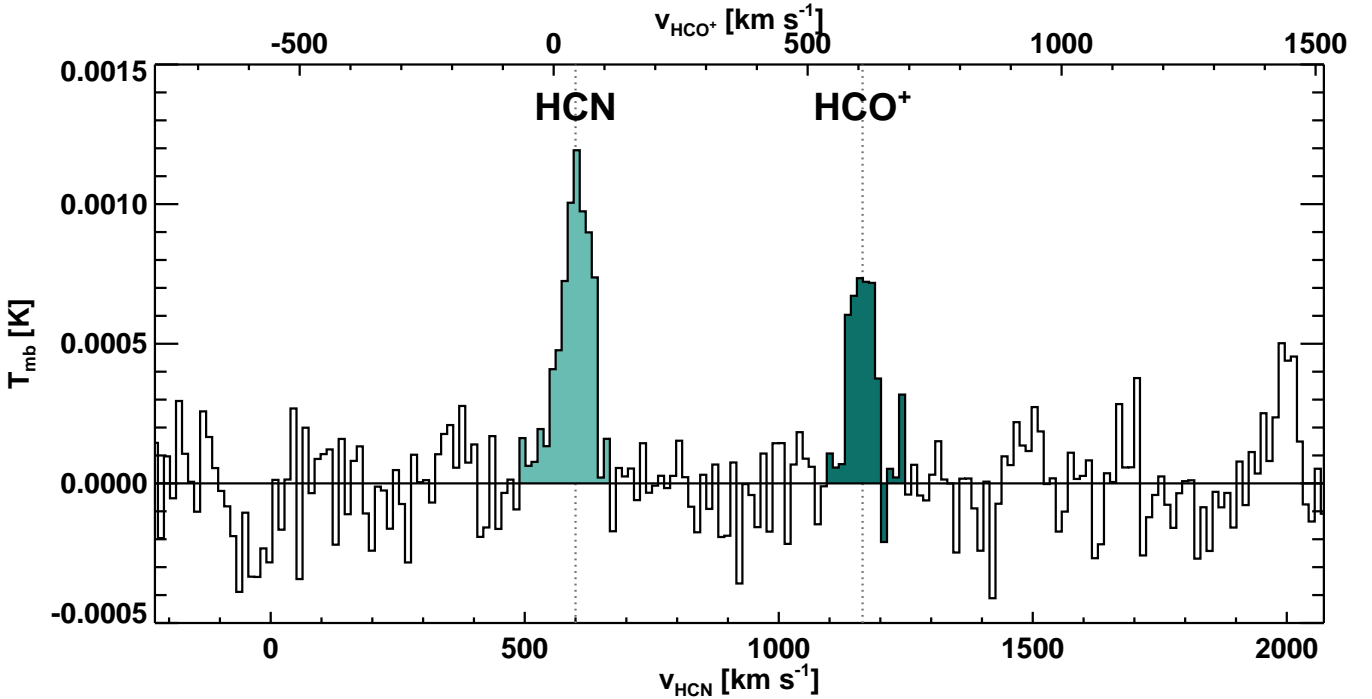}
\caption{The spectrum of HCN(1--0) and HCO$^+$ from OSO. The lines are robustly detected in the 43$''$ beam. These confirm the results of \citet{kohno+02} that the HCN is weak.}
\label{fig:oso_spec}
\end{figure*}

\begin{table}[t]
\caption{\bf Properties of NGC\,5195}
\begin{tabular}{lll}
\hline
{\bf RA (J2000)} & 13$^{\rm h}$29$^{\rm m}$59.59$^{\rm s}$ & (1)\\
{\bf Declination (J2000)} & +47$^\circ$15$^{\rm m}$58.1$^{\rm s}$ & (1)\\
{\bf Morphology} & I0 Peculiar & (2) \\
{\bf Distance} & 9.9\,Mpc & (3) \\
{\bf Position Angle} & +101$^\circ$ (CCW from N) & (2) \\
{\bf Inclination Angle} & 37$^\circ$ (face-on = 0$^\circ$) & (2) \\
{\bf Nuclear activity} & AGN\,/\,LINER & (4) \\
{\bf log({\em L}$_{\rm FIR}$)} & 9.3 $L_\odot$ & (5) \\
{\bf {\em M}$_\star$} & 1.95$\times$10$^{10}$\,M$_\odot$ & (6) \\
\hline
\end{tabular}
\label{tab:ngc5195}
\raggedright \\
(1) \citealt{2mass} (2) \citealt{devauc+91} (3) \citealt{tikhonov+09} (4) \citealt{moustakas+10} (5) \citealt{dale+09} (6) \citealt{lanz+13}
\end{table}

\begin{table}
\caption{\bf Derived molecular properties}
\begin{tabular}{lrrrr}
\hline \hline
 & $^{12}$CO(1--0) & $^{13}$CO(1--0) & CN(1--0) & CS(2--1) \\
\hline
$\nu_{\rm rest}$ \hfill [GHz] & 115.2712 & 110.2014 & 113.4910 & 97.9810 \\
$\theta_{\rm beam}$ \hfill [$''$] & 5.0$\times$4.4 & 7.4$\times$5.9 & 7.2$\times$5.8 & 8.3$\times$6.7 \\
K per Jy & 4.126 & 2.290 & 2.291 & 2.290 \\
$v_{\rm range}$ \hfill [km\,s$^{-1}$] & 463--799 & 448--729 & 298--740 & 473--674 \\
$\Delta v$ \hfill [km\,s$^{-1}$] & 10 & 20 & 40 & 50 \\
RMS \hfill [mJy\,bm$^{-1}$] & 12.0 & 3.3 & 4.7 & 4.0 \\
$I_{\rm peak}$ \hfill [K\,km\,s$^{-1}$] & 243.8$\pm$15.1 & 23.9$\pm$2.3 & 19.6$\pm$2.3 & 5.7$\pm$1.1\\
Mom0 area \hfill [$\Box''$]$^\ddagger$ & 1489 & 763 & 365 & 360 \\
\hfill [kpc$^2$] & 3.29 & 1.69 & 0.81 & 0.80\\
RMS/channel \hfill [mJy] & 97.3 & 13.0 & 13.2 & 9.6 \\
Line flux \hfill [Jy\,km\,s$^{-1}$] & 243.8$\pm$5.2 & 22.7$\pm$1.0 & 12.4$\pm$1.4 & 3.9$\pm$1.0 \\
S/N (Line) & 46.9 & 22.7 & 8.9 & 3.9\\
\hline
$L_{\rm line}$ \hfill [$L_\odot$] & 2859$\pm$61 & 255$\pm$11 & 143$\pm$16 & 39$\pm$10 \\
\hline \hline
\end{tabular}
\label{tab:mol_properties}
\raggedright
$^\ddagger$arcsec$^2$\\
\end{table}

\begin{figure*}
\includegraphics[width=0.99\textwidth]{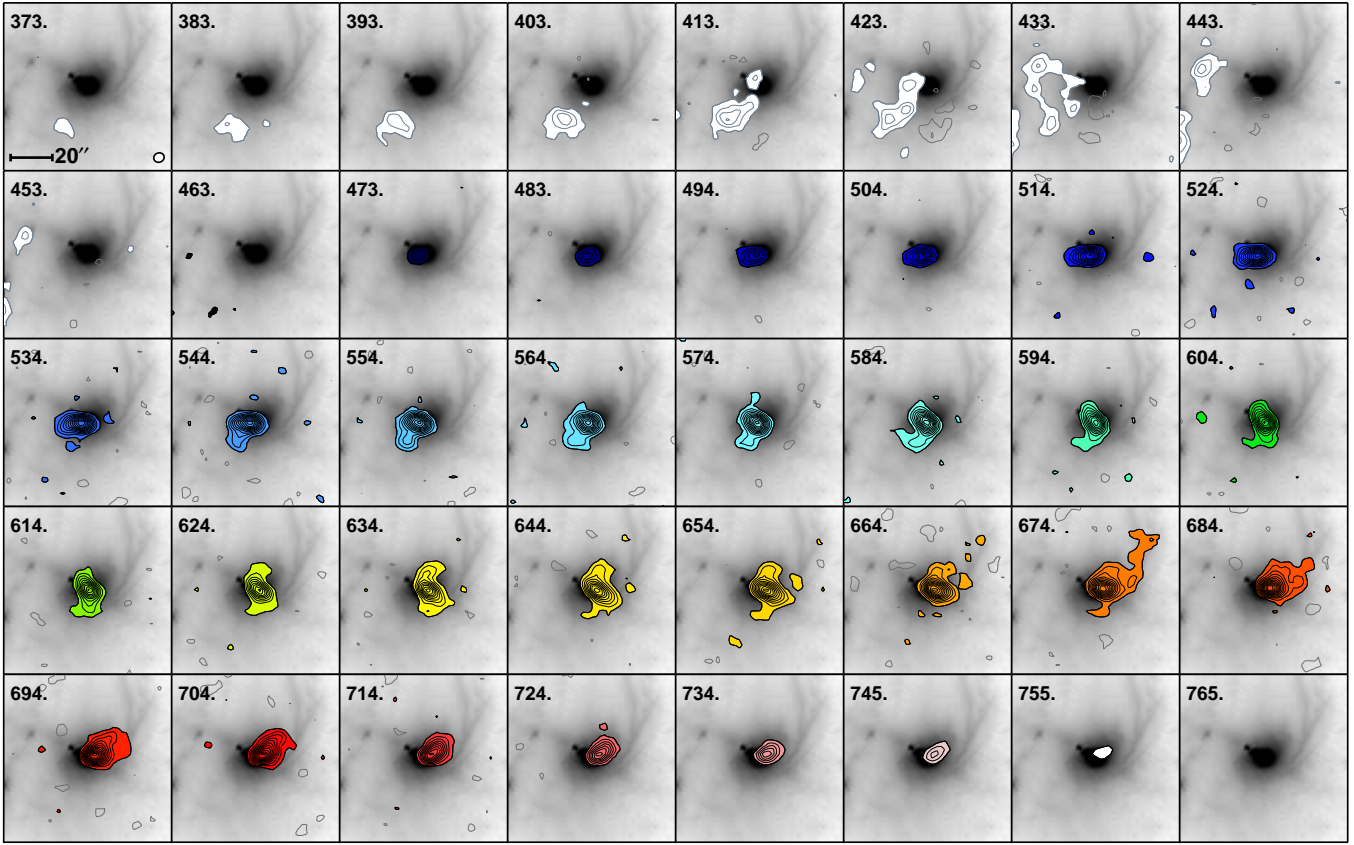} \vskip -1mm
\caption{\twco(1--0) channel maps of NGC\,5195 from CARMA (filled contours) overlaid on the nonstellar 8.0$\mu$m maps from {\em Spitzer} (grayscale), tracing the PAH emission. The contours are color coded based on the relative red- or blue-shift of the channel. The contours start at 3$\sigma$ (with $\sigma$ representing the RMS noise in Table~\ref{tab:mol_properties}), placed at 3$\sigma$ intervals. Negative contours (dark gray) are [-3,-6,-9]$\sigma$. Channels with \twco\ emission assumed to be associated with the M51a spiral arm contain a white fill and gray contours. The velocity color scale is identical to Fig.~\ref{fig:pah+co}. The velocities listed in the top left are optical recession velocities. The extension into the PAH filament can be seen between 634--694\,\kms.}
\label{fig:co_chans}
\end{figure*}
\begin{figure*}
\centering
\includegraphics[width=0.9\textwidth]{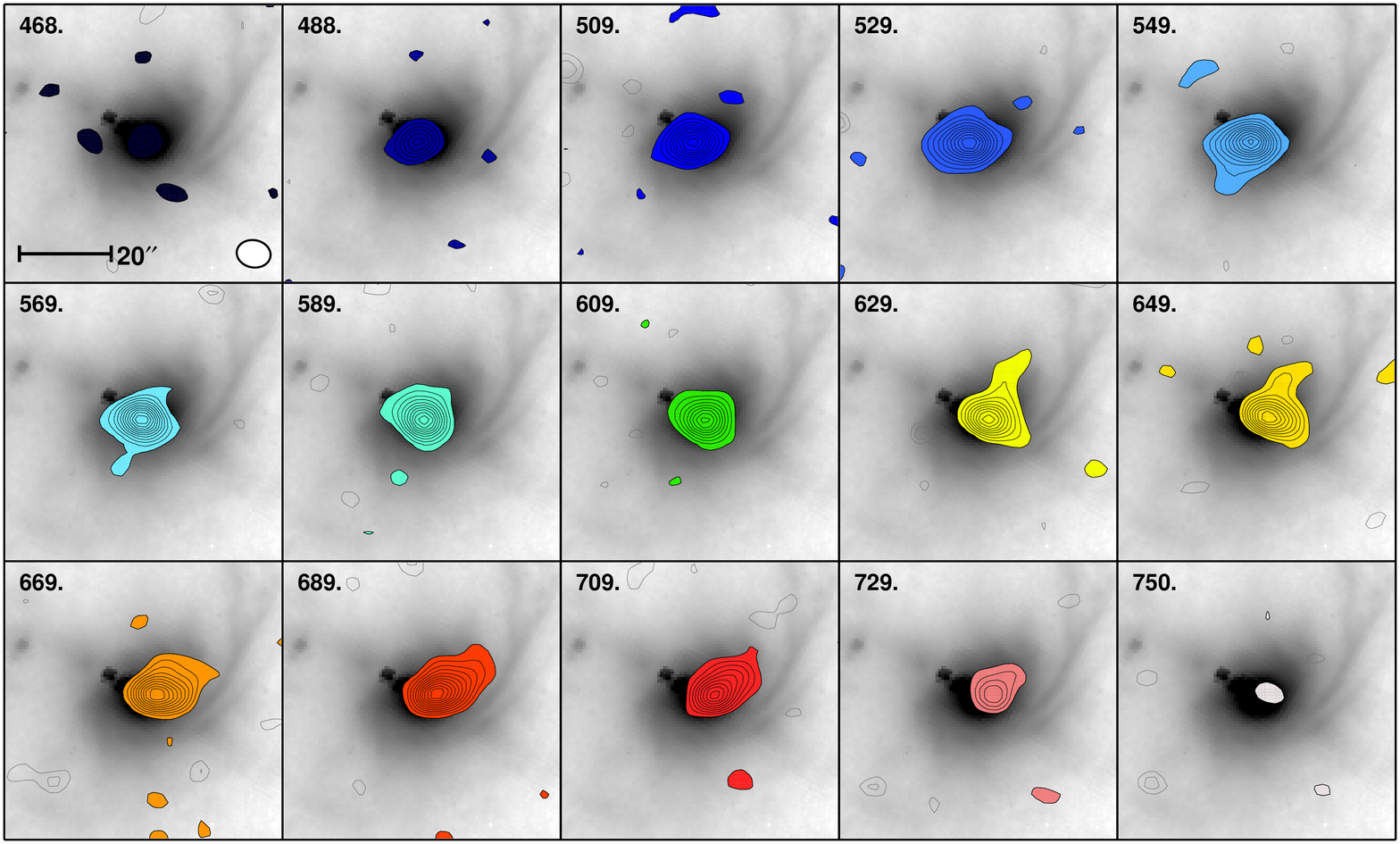} \vskip -1mm
\caption{\thco(1--0) channel maps of NGC\,5195 from CARMA (filled contours) overlaid on the nonstellar 8.0$\mu$m maps from {\em Spitzer} (grayscale), tracing the PAH emission. These contours are color coded based on the relative red- or blue-shift of the channel, placed at 2$\sigma$ intervals starting at 3$\sigma$. The velocities listed in the top left are optical recession velocities. The velocity color scale is identical to Fig.~\ref{fig:pah+co}.  A slight extension is seen along the PAH filament in \thco(1--0), but given the relative faintness of \thco\ compared to \twco, its likely that the more extended structure is below our detection threshold.}
\label{fig:co13_chans}
\end{figure*}

\begin{figure}
\includegraphics[width=0.49\textwidth]{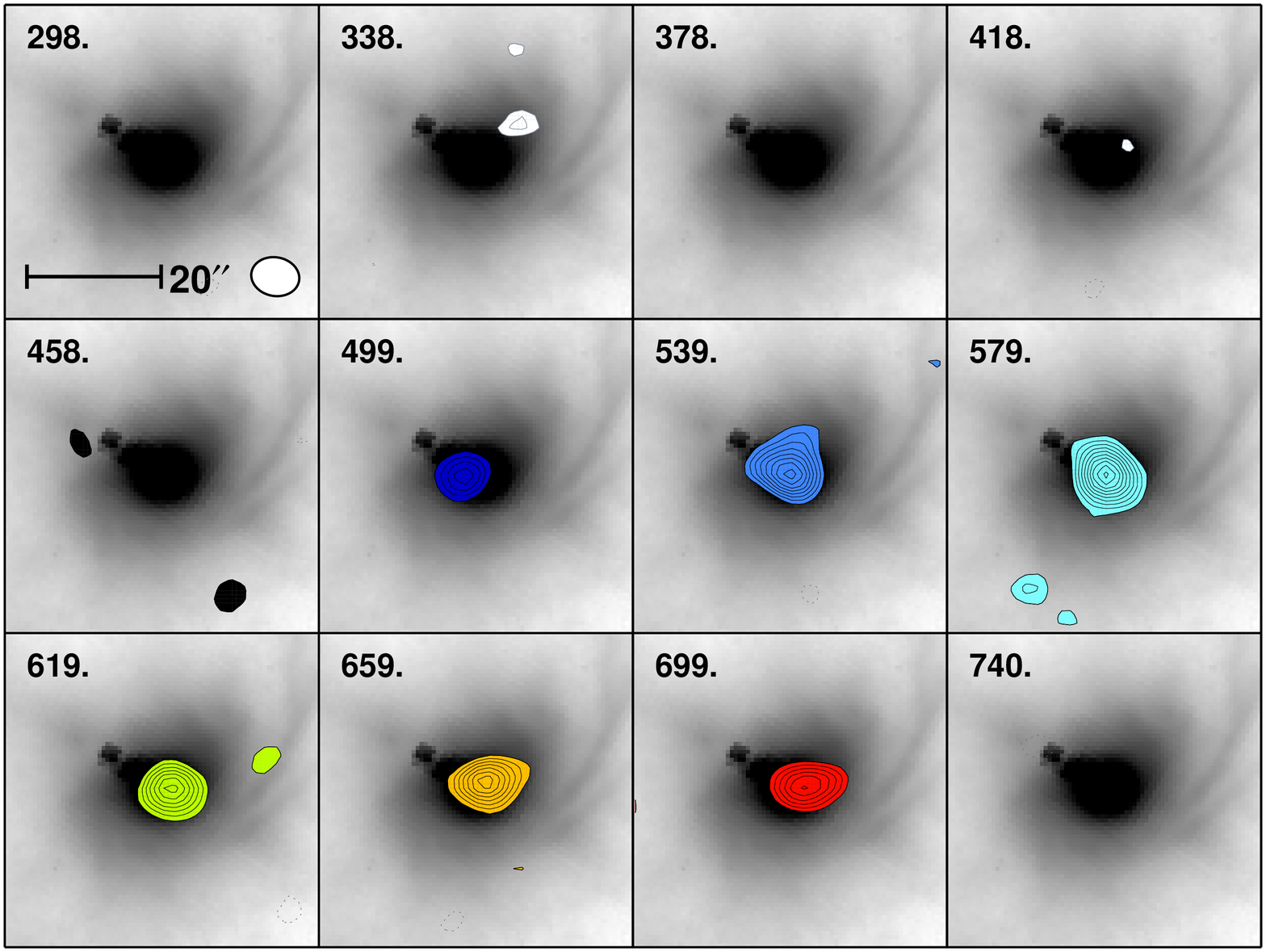} \vskip -1mm
\caption{CN(1--0) channel maps of NGC\,5195 from CARMA (filled contours) overlaid on the nonstellar 8.0$\mu$m maps from {\em Spitzer} (grayscale), tracing the PAH emission. The contours are color coded based on the relative red- or blue-shift of the channel, placed at 1$\sigma$ intervals starting at 3$\sigma$. The velocities listed in the top left are optical recession velocities. The velocity color scale is identical to Fig.~\ref{fig:pah+co}.  CN is well-detected in individual channels.}
\label{fig:cn_chans}
\end{figure}
\begin{figure}
\includegraphics[width=0.49\textwidth]{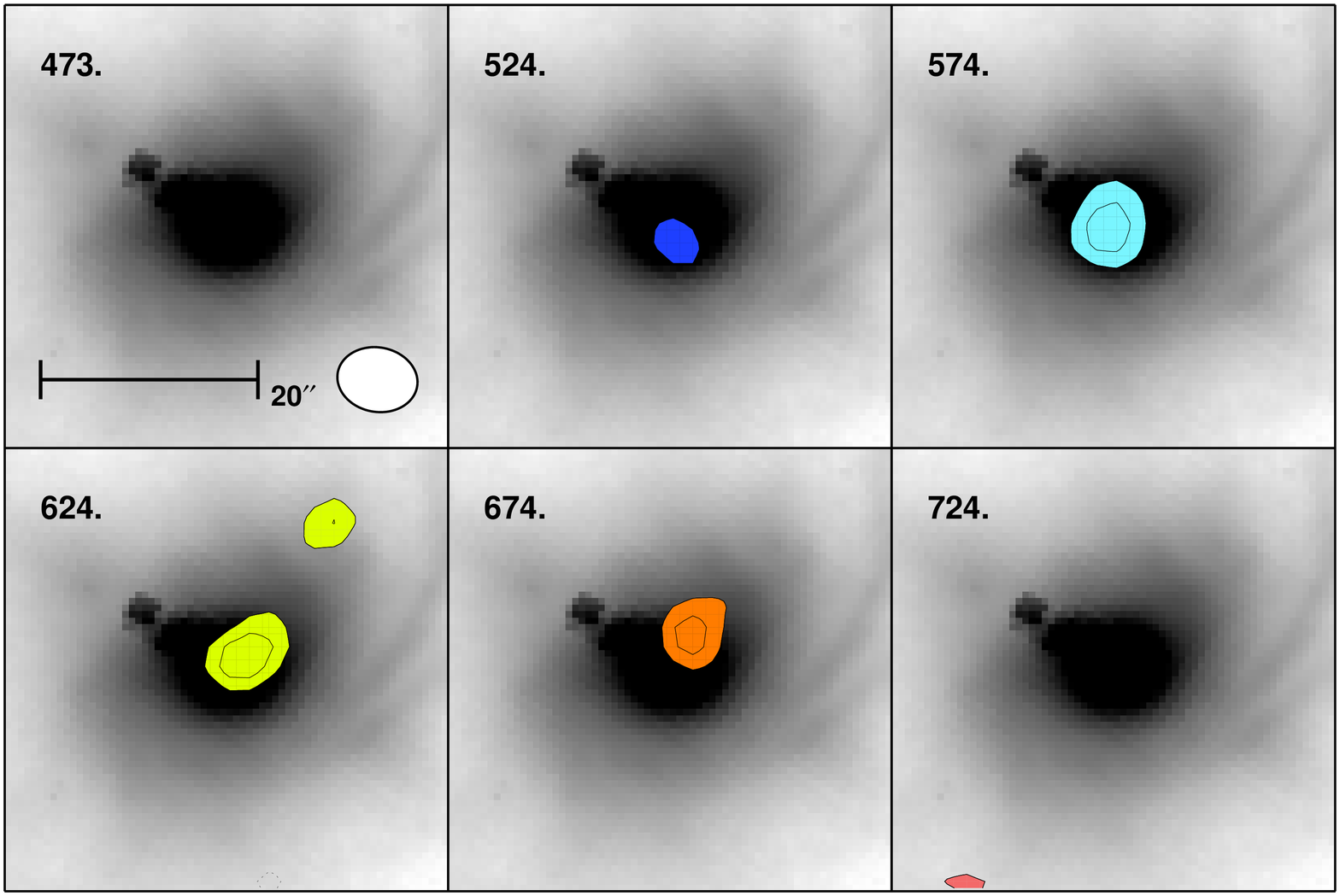} \vskip -1mm
\caption{CS(2--1) channel maps of NGC\,5195 from CARMA (filled contours) overlaid on the nonstellar 8.0$\mu$m maps from {\em Spitzer} (grayscale), tracing the PAH emission. The contours are color coded based on the relative red- or blue-shift of the channel, placed at 1$\sigma$ intervals starting at 3$\sigma$. The velocities listed in the top left are optical recession velocities. The velocity color scale is identical to Fig.~\ref{fig:pah+co}.  CS(2--1) is tentatively (S/N\,$\approx$\,3--4) detected in individual channels, but the total line is robustly detected.}
\label{fig:cs_chans}
\end{figure}

NGC\,5195 (alternatively known as M51b) is the disky barred (SB0/SBa(s)) galaxy \citep{spillar+92} that is the low mass companion of NGC\,5194 (alternatively known as M51a or ``the Whirlpool Galaxy''), having undergone a 3:1 interaction roughly 340--500\,Myr ago \citep{salo+00,dobbs+10,mentuch+12}. NGC\,5195 is a good case study of how star formation quenches for a galaxy a short time after an interaction, being nearby and showing signs of having recently transitioned. The integrated {\em u--r} color\footnote{Using the SDSS DR12 {\em u} and {\em r} modelmags \citep{sdssdr12}.} of NGC\,5195 is 2.25, which combined with its mass (Table~\ref{tab:ngc5195}), places it on the cusp between the red sequence  and the green valley \citep{schawinski+14}. Centrally concentrated molecular gas \citep{aalto5195}, and bright nuclear infrared emission \citep{boulade+96} further support that NGC\,5195 is in the process of becoming a poststarburst \citep{dressler+gunn83,zabludoff+96}, and finally an early-type red sequence galaxy, traversing the ``standard'' pathway for galaxy transitions \citep{hopkins+08}.

Table~\ref{tab:ngc5195} lists the basic properties of NGC\,5195 used for this paper. Mid-infrared (IR) spectra of NGC\,5195 confirm the presence of PAH emission \citep{boulade+96,roussel+07} that is more consistent with a poststarburst stellar population than an ongoing starburst. \citet{lanz+13} used the far-ultraviolet (UV) to the far-IR spectral energy distribution (SED) of NGC\,5195 and were able to estimate a modest star formation rate of 0.142\,$M_\odot$\,yr$^{-1}$. The warm H$_2$ emission detected with {\em Spitzer} Infrared Spectrograph is consistent with that of other star-forming galaxies in the SINGS sample \citep{roussel+07}. \citet{mentuch+12} also fit the ultraviolet to far-IR photometry and concluded that NGC\,5195 is a poststarburst. Though \citet{schweizer77} detected ionized gas emission lines in the galaxy, \citet{ho+97} and \citet{moustakas+10} found that the ionized gas line ratios are consistent with a low ionization nuclear emission line region (LINER; \citealt{kewley+06}). Weak X-ray emission was found coincident with a radio source, possibly due to an AGN, but also consistent with an ultra-luminous X-ray source origin \citep{terashima+04,schlegel+16}.

\citet{sage90} mapped NGC\,5195 in \twco(1--0), and observed \twco(2--1) and \thco(1--0) with the National Radio Astronomy Observatory (NRAO) 12m single dish, showing that there was molecular gas in the galaxy. \citet{kohno+02} were able to detect HCN(1--0) using Nobeyama Radio Observatory (NRO) 45m single dish, and mapped \twco(1--0) using the Nobeyama Millimeter Array (NMA). \citet{kohno+02} showed that the molecular gas is concentrated in the center of NGC\,5195, and claimed that the molecular gas is too stable to form stars.

\begin{figure*}[t]
\includegraphics[width=0.99\textwidth]{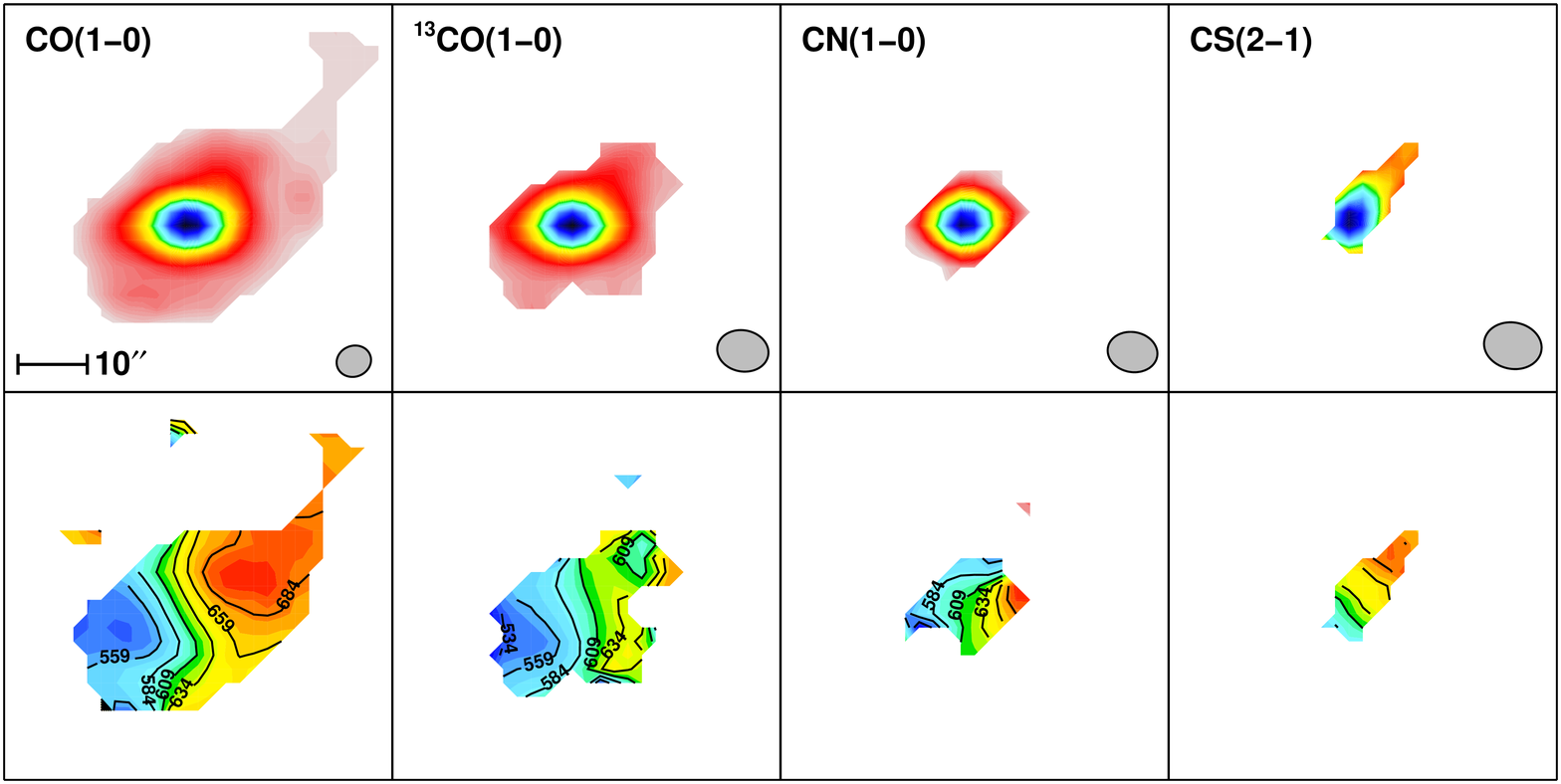}
\caption{Moment maps of all detected molecules, including \twco(1--0), \thco(1--0), CN(1--0), and CS(2--1). The top row displays the clipped integrated intensity maps of each molecular species detected, and the bottom row displays the mean velocity maps of each species. The velocity color scale is identical to Fig.~\ref{fig:pah+co}. The CS(2--1), the faintest line, traces only the core of the molecular gas in NGC\,5195, which supports the idea that the central region contains the densest gas. The \twco, \thco, and CS trace similar velocity structures, whereas the velocity map of the CN seems to trace different kinematics.}
\label{fig:moments}
\end{figure*}

We present new molecular gas maps of NGC\,5195, and use them to develop a narrative of the star formation since the time of its interaction with M51a.  In \S\ref{sec:obs}, we describe the observations from CARMA, including reduction and analysis methods.  In \S\ref{sec:molgas}, we discuss the properties of the molecular gas of NGC\,5195. In \S\ref{sec:sf} we discuss the past and present star formation in NGC\,5195.  In \S\ref{sec:conc}, we present our conclusions.  We use the common distance to M51a and NGC\,5195 of 9.9\,Mpc, following \citet{tikhonov+09} who used the Hertzsprung-Russell diagram of individual resolved stars within the M51 system to calculate the distance. The corresponding spatial scale is 47 parsecs per arcsecond.

\section{Observations and Analysis}
\label{sec:obs}
\subsection{Ancillary data}
\label{sec:ancillary}
The 3.6--4.5--8.0\micron\ {\em Spitzer} \citep{spitzer} Infrared Array Camera (IRAC; \citealt{irac}) data that are shown in Figure~\ref{fig:pah+co} were obtained by downloading supermosaics from the {\em Spitzer} Heritage Archive\footnote{\href{http://sha.ipac.caltech.edu/applications/Spitzer/SHA/}{http://sha.ipac.caltech.edu/applications/Spitzer/SHA/}}. These data were originally part of the {\em Spitzer} Infrared Nearby Galaxy Survey \citep{sings}. The observations were taken on May 2004, with exposure times of 26.8 seconds and were processed through the {\em Spitzer} Science Center reduction pipeline, version S18.25.

The {\em Herschel} Photo-detecting Array Camera and Spectrometer (PACS; \citealt{pacs}) data were observed on December 2009 in the ``blue channel'' (70$\micron$ and 160$\micron$) and were downloaded from the Herschel Science Archive (PI: C. Wilson, KPGT\_cwilso01; see: \citealt{parkin+13} for the 70\micron\ image). The Level-1 data were reduced using {\sc scanamorphos} \citep{scanomorphos} package that fits accurately the low-frequency (1/$f$) noise. Typical sky background measurements in the final maps were 0.13 and 0.25 mJy for the 70 and 160$\micron$ bands, respectively.

The optical spectrum used below was obtained through the Sloan Digital Sky Survey (SDSS), which surveyed galaxies using a dedicated 2.5m telescope at Apache Point Observatory. The reduced and calibrated spectral data were obtained from SDSS Data Release 12 \citep{sdssdr12}, without any additional modifications applied.

The 1.4\,GHz data for NGC\,5195, taken as part of the Westerbork SINGS survey \citep{braun+07} was downloaded for analysis. Observations of the M51 complex were taken on dates June 25, 2003 and Nov 23, 2003, with a beamsize of 17$\times$12.5$''$. Because the 1.4\,GHz emission associated with NGC\,5195 has a complex distribution, we measured the 1.4 GHz flux density from the SINGs image by summing the flux within an aperture matching the extent of the CO emission using the Common Astronomy Software Application ({\sc casa})\footnote{\href{https://casa.nrao.edu/}{https://casa.nrao.edu/}} viewer \citep{casa}.

\subsection{OSO 20m observations}	
HCN(1--0) and HCO$^+$(1--0) were observed with the Onsala Space Observatory, Sweden, on 3--4 March, 15, and 26 April 2016. The SIS 3-mm receiver was connected to the Omnisys A spectrometer, which gives a bandwidth of 4\,GHz wide with dual polarization. The observations were performed with a dual beam switch mode with a throw of 11$'$ in azimuth. The half power beam width of the telescope at the rest frequency (88.9 GHz) was 43$''$ and the pointing accuracy was better than 3$''$. The focus was checked each day on bright quasars and the corrections were $\sim$1\,mm. The average system temperature was around 150\,K, and the opacity 0.2. 

We converted the data from antenna temperature to main beam temperature  correcting by a beam efficiency of 0.53.  The  spectrum was averaged to a final velocity resolution of 40\,km\,s$^{-1}$ (12 MHz), for which the achieved RMS is 0.145\,mK. We fitted a baseline of order two to subtract the continuum. 

The spectra of the HCN and HCO$^+$ lines is shown in Fig.~\ref{fig:oso_spec}. Using the K per Jy factor of 0.07999, we measure a total HCN(1--0) line flux of 3.25$\pm$0.27\,Jy\,km~s$^{-2}$, with velocity width of 219\,km\,s$^{-1}$. We also measure a HCO$^+$(1--0) total line flux of 2.09$\pm$0.24\,Jy\,km~s$^{-2}$ with a velocity width of 193\,km\,s$^{-1}$.

We measure a HCN/HCO$^+$ J=1--0 line ratio of 2. Reproducing HCN(1--0)/HCO$^+$(1--0) of 2 with $n$\,=\,10$^4$\,cm$^{-3}$ and T$_K$\,=\,65\,K requires a higher HCN than HCO$^+$ abundance. As long as the emission is not very optically thick then the lower critical density of HCO$^+$ would render its J=1--0 transition more luminous than that of HCN  (for the same abundance). For the given physical conditions above the HCN(1--0)/HCO$^+$(1--0) ratio should be about 0.5 for the same abundance (and for moderate optical depths). To obtain HCN(1--0)/HCO$^+$(1--0) of 2 for $n$\,=\,10$^4$\,cm$^{-3}$ and T$_K$\,=\,65\,K requires the HCN/HCO$^+$ abundance ratio to be about 10.

\subsection{CARMA Observations}
We present observations of NGC\,5195 that were both pulled from the CARMA archive, as well as original observations. The archival portion of our NGC\,5195 data was observed in \twco(1--0) with CARMA between 2012 May and 2012 December in two different configurations: C-array (1$''$, 30--350m baselines), D array (3$''$, 11--150m baselines) by the CArma and NObeyama Nearby galaxies (CANON; \citealt{donovanmeyer+13}) survey. We followed up these archival observations in 2014 June with observations of NGC\,5195 in E-array (8$''$, 8--66m baselines), to better detect some of the more diffuse gas that might be present.  \twco(1--0),\thco(1--0), CN(1$_{0,2}$--0$_{0,1}$) and CS(2--1) were observed simultaneously in this set of E-array observations, utilizing the upgraded correlator. The observing properties for each molecular line detected by CARMA in NGC\,5195 are listed in Table~\ref{tab:mol_properties}. All data were reduced and imaged using the Multichannel Image Reconstruction, Interactive Analysis and Display software ({\sc miriad}; \citealt{miriad}).

The primary beam has a diameter of 2$'$ at CO(1--0), which covers all emission from NGC\,5195. M51a itself does appear within the primary beam, but is  easily separable from NGC\,5195 (see Fig.~\ref{fig:pah+co}a).  All observations used a long integration on a bright quasar to calibrate the passband, and alternated integrations between a gain calibrator \hbox{(J1419+543)} and NGC\,5195.  We used the {\sc miriad} task {\tt uvlin} to separate out continuum emission from the line emission, and we estimate the continuum flux of NGC\,5195 to be $1.21\pm0.27$\,mJy, centered on 106\,GHz, from a CARMA image with a 7.6$''\times$6.1$''$ beam.\footnote{This does not include the absolute flux calibration uncertainty of 20\%.} Calibration and data reduction steps were followed identically to those in \citet{alatalo+13} to construct channel maps, and moment maps.

Channel maps are shown in Figures~\ref{fig:co_chans}--\ref{fig:cs_chans}, with the individual gas channels overlaid on a 8$\mu$m nonstellar {\em Spitzer} IRAC image. To create the underlying {\em Spitzer} image, the 8.0$\mu$m IRAC4 data was corrected for the stellar contribution by subtracting the 3.6$\mu$m IRAC1 image, normalized to the expected stellar continuum at 8$\mu$m (0.232; \citealt{helou+04}). \twco, \thco, and CN(1--0) are detected.  The CN(1--0) channel map includes blueshifted emission compared to the line peak ($\approx$\,250\,km~s$^{-1}$ away, at 340\,km~s$^{-1}$), which we have attributed to M51a and do not consider in subsequent analysis. CS(2--1) is weak, but still detected with a signal-to-noise (S/N) of 3.9. A tentative line, identified as CH$_3$CN(6$_k$--5$_k$), seems to be present in the \thco\ spectrum outside of the velocity range that would be inhabited by \thco\ in M51a, and has a line flux of 1.27$\pm$0.52 (only detected with 2.4$\sigma$ significance), but requires deeper observations to confirm whether it is real.

\begin{figure}
\includegraphics[width=0.49\textwidth,clip,trim=0cm 0cm 0cm 0cm]{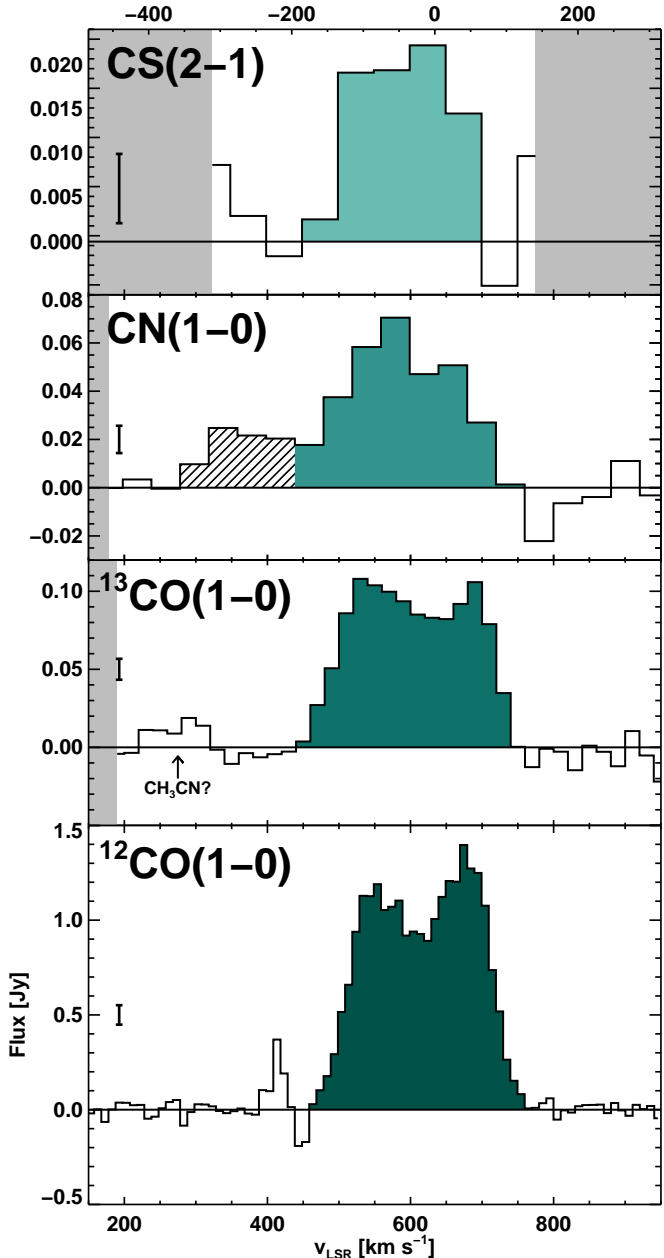} \vskip -2mm
\caption{The extracted spectra of the CARMA-detected molecular lines. The moment0 maps were used as masks to integrate the data cubes to create the spectra of CS(2--1) (top), CN(1--0) (top middle), \thco(1--0) (bottom middle) and \twco(1--0) (bottom). The position of the tentatively identified CH$_3$CN is shown in the \thco(1--0) spectrum. The error bar at the left of each panel marks the 1$\sigma$ noise in the spectrum. The shaded gray regions represent the edges of the correlator bands used for each of the spectra. The CN(1--0) spectrum shows emission that does not appear to be associated with the CN(1--0) line in NGC\,5195 (see Fig.~\ref{fig:cn_chans}), and is shaded with diagonal lines.}
\label{fig:spectra}
\end{figure}

\begin{figure*}
\centering
\subfigure{\includegraphics[width=0.7\textwidth]{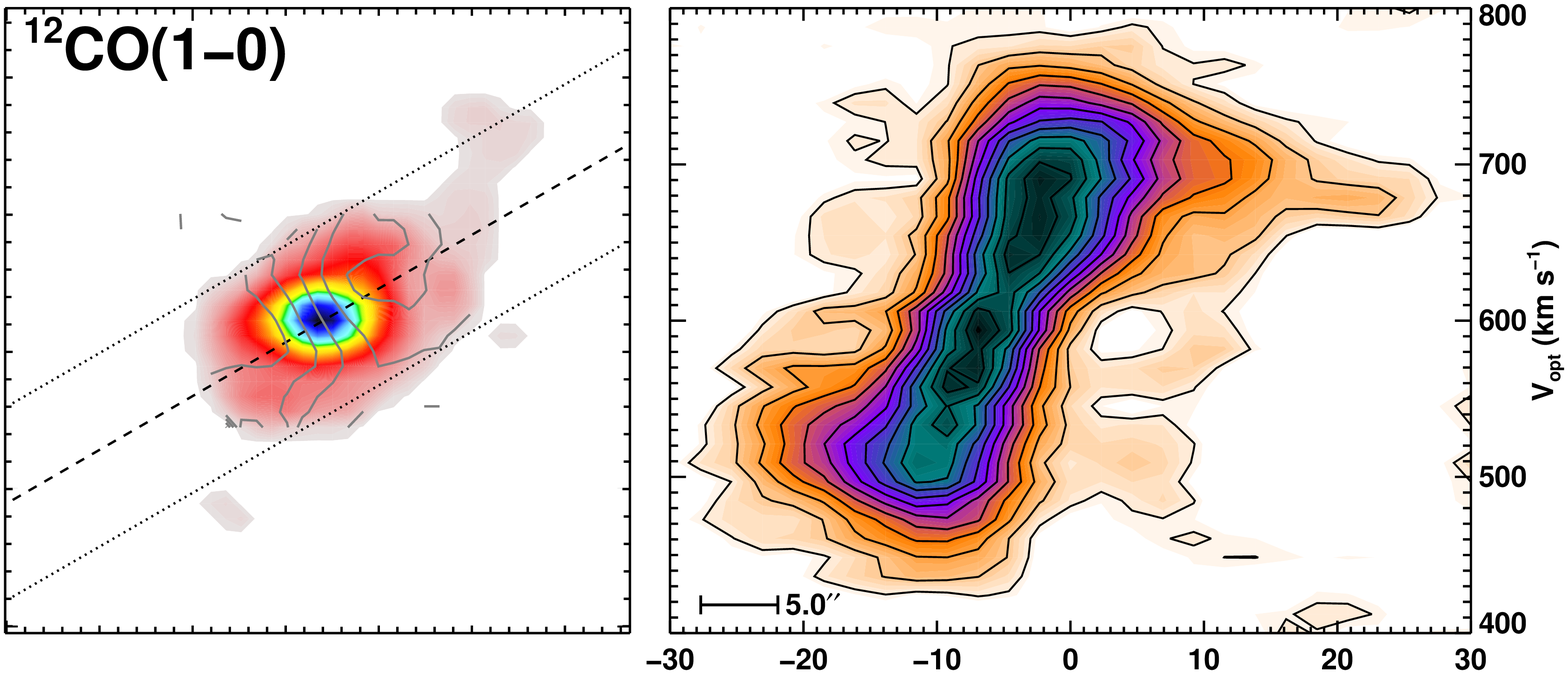}} \vskip -3mm
\subfigure{\includegraphics[width=0.7\textwidth]{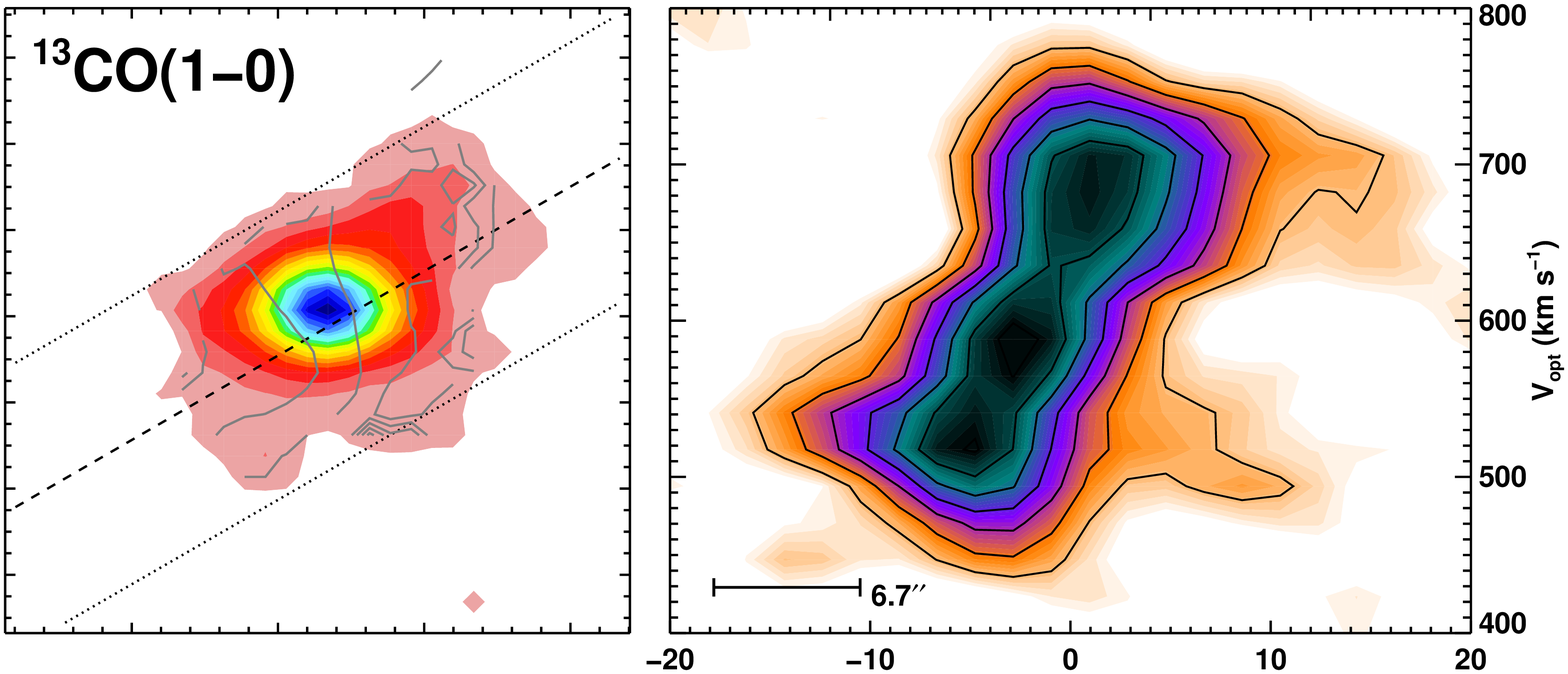}} \vskip -3mm
\subfigure{\includegraphics[width=0.7\textwidth]{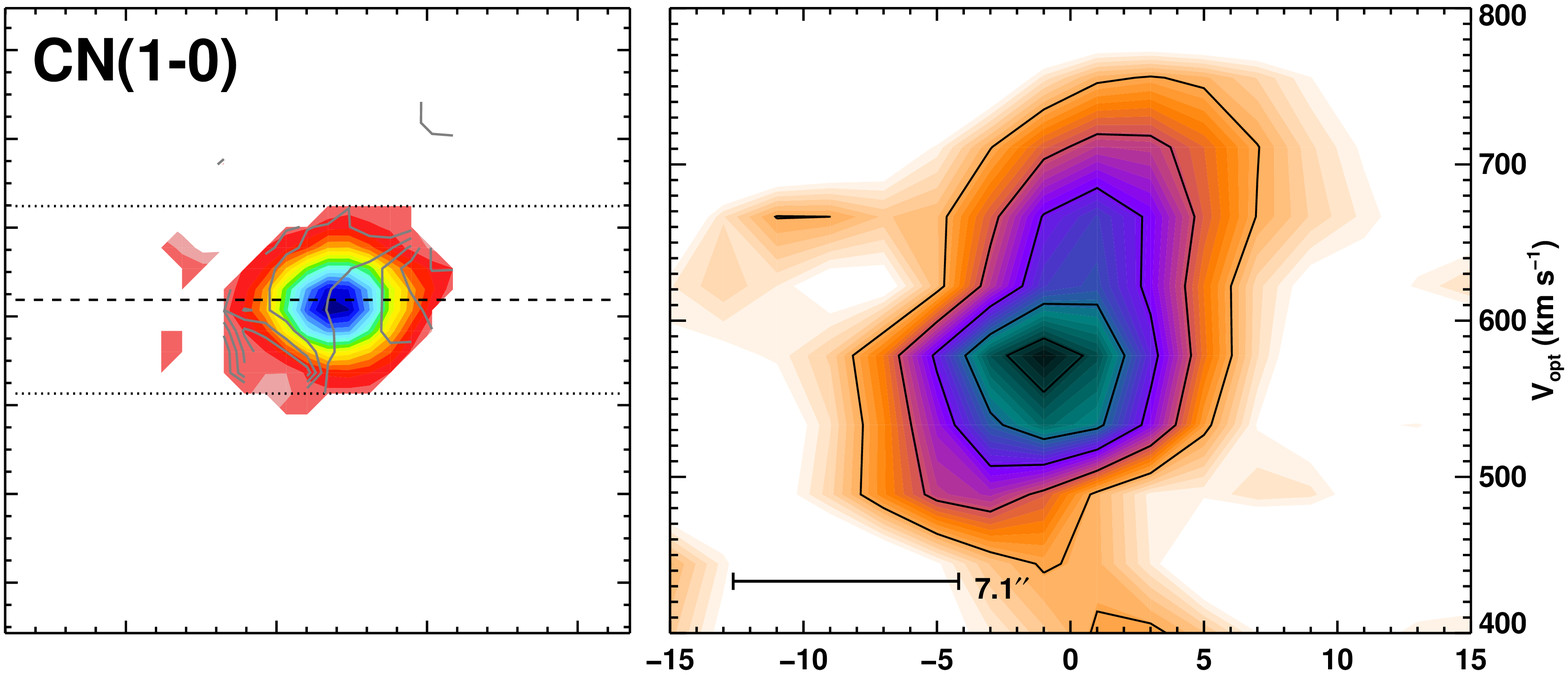}} \vskip -3.5mm
\caption{The position-velocity (PV) diagrams are shown for \twco(1--0) (top), \thco(1--0) (middle) and CN(1--0) (bottom). The left-hand images show the moment0 map of each molecular species. The dotted line denotes the bisector at the chosen position angle across the cube to create the PV diagram (right). The dot-dashed lines denotes the upper and lower bounds for the sum for each sliced region. Both \twco\ and \thco\ share a common position angle, and show signs of regular rotation. The CN required a different position angle, possibly suggesting an origin distinct from the other molecular tracers. The CN could also be tracing a different (denser) gas phase than both \twco\ and \thco\ (see \S\ref{sec:cn}).}
\label{fig:pvdiags}
\end{figure*}

Figures~\ref{fig:pah+co} and \ref{fig:moments} display the integrated intensity (moment0), and mean velocity (moment1) maps, and Figure~\ref{fig:spectra} shows the integrated spectra of the \twco, \thco, CN(1--0), and CS(2--1) within NGC\,5195, Figure~\ref{fig:pvdiags} shows the position-velocity diagrams across the specified slices of the \twco, \thco, and CN detections (the CS(2--1) being too low in S/N and compact to display). Of note, while the \twco\ and \thco\ share the same kinematic axis (of 30$^\circ$), the CN (with an angle of 0$^\circ$) has a different kinematic major axis from the other gas tracers and is thus considered kinematically misaligned according to \citet{davis+11}.

The integrated spectrum for each line was constructed using the moment0 map to create a clip-mask and integrating the flux within the moment0-defined (unmasked) aperture in each channel of the data cube. This was done separately for each tracer. The root mean square (RMS) noise was then calculated as the standard deviation of all pixels in the cube outside of the moment0-aperture per channel and is listed in Table~\ref{tab:mol_properties}.  An additional 30\% correction was also added in quadrature to the RMS noise to account for the oversampling of the maps (see: \citealt{a15_co13} for details).  The RMS noise per channel for the spectrum was then calculated by multiplying the RMS of the entire data cube by the square root of the total number of beams in the moment0-aperture.

The line luminosity for each molecule was calculated using \citet{solomon+05}: 
\begin{equation}
L_{\rm line} = 1.04\times10^{-3}~S_{\rm line}\Delta v~\nu_{\rm rest} (1+z)^{-1} D_L^2
\end{equation}
\noindent where $S_{\rm line}\Delta v$ is the line flux in [Jy\,km\,s$^{-1}$], $\nu_{\rm rest}$ is the rest frequency of the line in [GHz], {\em z} is the redshift of NGC\,5195 (calculated using a systemic velocity of 634~km\,s$^{-1}$, the \hbox{CO(1--0)} spectrum), and $D_L$ is the luminosity distance, defined as 9.9\,Mpc in \S\ref{sec:intro}. The resultant line fluxes are listed in solar luminosities in Table~\ref{tab:mol_properties}.

\subsection{Comparison to Previous Detections}
\label{sec:comparison}
\citet{kohno+02} observed \twco(1--0) and HCN(1--0) in NGC\,5195 with the NRO 45m single dish, and imaged the \twco(1--0) with the NMA Interferometer. The new OSO HCN(1--0) observations are consistent with the relatively faint HCN emission found by \citet{kohno+02}. Using the 45m, \citet{kohno+02} find a total line flux of 199~Jy\,km\,s$^{-1}$ (using a 2.4 Jy per K conversion; \citealt{ueda+14}), in conflict with their resultant interferometric flux of 340~Jy\,km\,s$^{-1}$. We have measured a \twco(1--0) line flux of 243.8$\pm$5.2~Jy\,km\,s$^{-1}$, falling right between the two estimates of those authors (34\% larger than the 45m, and 27\% lower than the NMA line fluxes). Were some of the M51a emission to be included in the map, it would result in an overestimation of the CO(1--0) line flux for NGC\,5195, though \citet{kohno+02} make note of the low velocity component likely being from the spiral arm of NGC\,5194 (displayed as gray contours in the first 9 channels in Figure~\ref{fig:co_chans}). If we also account for the 20\% absolute flux uncertainties associated with the 3mm flux calibration (for both the NMA and the CARMA observations), we consider ourselves to be in reasonable agreement with \citet{kohno+02}.

\citet{sage90} presented \twco(2--1) and \thco(1--0) observations of NGC\,5195 from the NRAO~12m, measuring a \thco\ line flux of 12.5$\pm$3.3~Jy\,km\,s$^{-1}$ (using the Jy/K\,=\, 30.4 of the NRAO~12m; \citealt{ueda+14}). We have detected nearly twice the total \thco\ flux of \citet{sage90}, though that is likely due to signal-to-noise (our detection has six times higher signal-to-noise than \citealt{sage90}). It is also possible that pointing errors and calibration uncertainties might also play a part. \citet{matsushita+10} used the NRO~45m to measure the \thco(1--0) of NGC\,5195 to be 28.8~Jy\,km\,s$^{-1}$, in much better agreement (within 22\%) with our \thco\ measurement. Overall, our observations are within reasonable agreement with those done previously. 

\subsection{Molecular mass and H$_2$ surface density in NGC\,5195}
We calculate the molecular mass expected in NGC\,5195 using the \twco(1--0) luminosity. Assuming that NGC\,5195 is similar to the Milky Way, where we assume that molecular gas is distributed within giant molecular clouds (GMCs), the conversion between the CO luminosity and the H$_2$ mass (from \citealt{bolatto+13}) is:
\begin{equation}
\frac{M_{\rm H_2}}{L_{\rm CO}} = 8.75\times10^4~\frac{M_\odot}{L_\odot}
\end{equation}
\noindent which corresponds to a $X_{\rm CO}$ conversion factor of 2$\times$10$^{20}$\,cm$^{-2}$\,(K~km\,s$^{-1}$)$^{-1}$, considered the standard, holding both for the Milky Way and most normal nearby galaxies. Using this conversion, and $L_{\rm CO}$\footnote{The CO line luminosity} from Table~\ref{tab:mol_properties}, we find a total H$_2$ mass of (2.74$\pm0.06)\times$10$^8$~$M_\odot$\footnote{Does not include the 30\% mass conversion uncertainty \citep{bolatto+13} or 20\% absolute flux uncertainty}.

To calculate the H$_2$ surface density, $\Sigma_{\rm H_2}$, we use the area subtended by the \twco(1--0) clipped moment map (listed in Table~\ref{tab:mol_properties}), converted from $\Box''$ to the physical scale for the galaxy to be 3.15\,kpc$^2$, which corresponds to a mean H$_2$ surface density, $\langle\Sigma_{\rm H_2}\rangle = 87$$\pm$$2$\,$M_\odot$\,pc$^{-2}$. To calculate the peak $\Sigma_{\rm H_2}$, we convert the peak line intensity to the H$_2$ surface density, assuming the conversion from \citet{bolatto+13}:
\begin{equation}
\label{eqn:sigh2}
\frac{\Sigma_{\rm H_2}}{M_\odot\,{\rm pc}^{-2}} = \frac{3.16\,I_{\rm peak}}{\rm K~km~s^{-1}}
\end{equation}
\noindent where $I_{\rm peak}$ represents the peak intensity of the \twco\ map, resulting in $\Sigma_{\rm H_2,peak}$\,=\,810$\pm$50\,$M_\odot$\,pc$^{-2}$, made over an area of 0.05\,kpc$^2$ (one beam), an order of magnitude larger than $\langle\Sigma_{\rm H_2}\rangle$, and consistent with the range seen in nearby disc galaxies \citep{bigiel+08,leroy+08}.

\begin{figure*}
\includegraphics[width=0.99\textwidth,clip,trim=0cm 0cm 0cm 0cm]{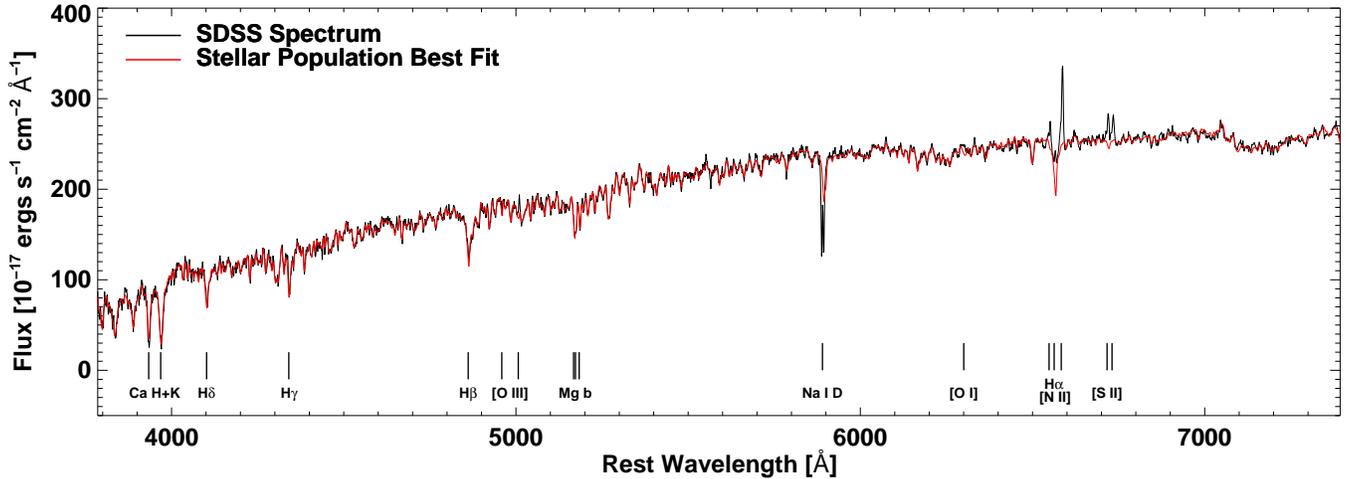} \vskip -2mm
\caption{The nuclear spectrum of NGC\,5195 (black) overlaid with the best fit combination of stellar models (red) from {\sc miles} \citep{miles} and {\sc ppxf} \citep{ppxf}. Optical lines of note are labeled. The stellar models that best fit these data were a combination of intermediate age ($\sim$\,1\,Gyr) and older ($\sim$\,10\,Gyr) stellar populations (shown in Fig.~\ref{fig:stelpop}).}
\label{fig:stelmodels}
\end{figure*}

\begin{figure}
\includegraphics[width=0.49\textwidth,clip,trim=0cm 0cm 0cm 0cm]{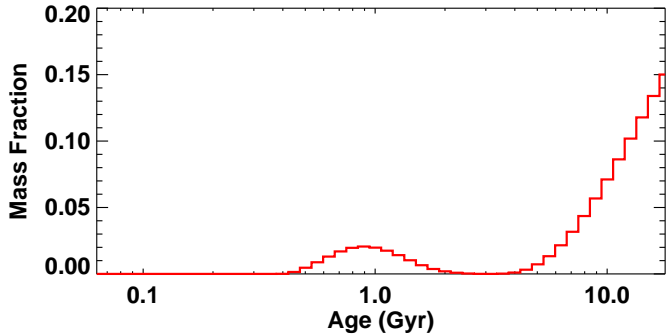} \vskip -1.5mm
\caption{The probability distribution function of the stellar populations in NGC\,5195 fit to the central spectrum from SDSS DR12 \citep{sdssdr12}. We used {\sc ppxf} \citep{ppxf} to fit a smoothed star formation history using the {\sc miles} library \citep{miles} with a Salpeter IMF. NGC\,5195 is primarily composed of older ($\gtrsim$\,10\,Gyr) stars, but has a significant fraction ($\approx$\,20\%) of intermediate age stars ($\approx$\,1\,Gyr).}
\label{fig:stelpop}
\end{figure}


\section{The state of the molecular gas in NGC\,5195}
\label{sec:molgas}

\subsection{The revised \thco/\twco\ ratio}
\citet{sage90} measured the \twco\ and \thco\ in NGC\,5195 (discussed in \S\ref{sec:comparison}), finding a \twco/\thco\ flux ratio (hereafter, \rco) to be $\approx$26. This high ratio is often seen in hot dust hosts and ULIRGs \citep{aalto+95}, suggestive of a hot, disrupted molecular disk in which the \twco\ emission includes a large optically thin contribution. The CARMA-derived \twco\ and \thco\ fluxes shown in Table~\ref{tab:mol_properties} update \rco\ in NGC\,5195 to be 11.4$\pm$0.5, which is more consistent with normal star-forming galaxies as well as field early-type galaxies \citep{a15_co13} than interacting galaxies. The revised \rco\ in NGC\,5195 seems to suggest that the dynamical state of the gas is not as disrupted as originally thought, having had time to settle since the last interaction.

\subsection{What is CN tracing?}
\label{sec:cn}
The CN abundance can be enhanced either by UV fields (for example as a photo-dissociation product of HCN in the envelopes of molecular clouds) or by X-rays, that can penetrate further in the inner,  denser regions. Hence, CN is a molecular tracer of photo-dissociated regions (PDRs) and X-ray dissociated regions (XDRs) depending on the environment (e.g. \citealt{lepp+96,rodriguez-franco+98}). The strong PAH, IR, and H$\alpha$ emission in NGC\,5195 (\citealt{boulade+96,greenawalt+98} and Fig.~\ref{fig:pah+co}) suggest that the CN is associated with the old starburst caused by the encounter with M51a and may be associated with PDRs.

To explore the possibility that CN could be tracing XDRs, we compared our CN intensity maps with the X-ray emission from \citet{schlegel+16}, where {\em Chandra} images show two recent outbursts coming out from the supermassive black hole in NGC\,5195. A comparison of the CN and X-ray images shows no obvious spatial correlation between them.


In a barred galaxy potential, dynamical models predict 2 kinds of orbits: 1) {\em x}1 orbits oriented parallel to the major axis of the bar that support bar and 2) {\em x}2 orbits aligned perpendicular to the bar and located closer to a galaxyÕs nucleus.  Gas in self-intersecting {\em x}1 orbits tends to shock and lose energy, eventually changing orientation and moving inward to the lower-energy {\em x}2 orbits.  For a review, we refer readers to \citet{alloin06}. The moment maps in Fig.~\ref{fig:moments} show that CN arises from the inner regions of the galaxy with isovelocity contours at a different angle than those of \twco\ and \thco. The latter seem to trace a more extended gas distribution, possibly associated with the {\em x}1 orbits of the large-scale bar \citep{kohno+02}, while CN could be tracing the perpendicular {\em x}2 orbits of the inner region. It is possible that this difference in alignment is due to resolution effects, but given the high signal-to-noise CN channels seen in Fig.\,\ref{fig:cn_chans}, we believe the detection of differing alignments is robust.

This would be a similar scenario to that observed in the central 100\,pc of M\,82, where the CN abundance (among other PDR tracers) is enhanced by a factor of 3 in the {\em x}2 orbits \citep{ginard+15}.  This result is consistent with NGC\,5195 being a post-burst system, with forbidden neon lines present in the IR spectrum \citep{boulade+96} and the optical lines (H$\alpha$ absorption and strong [N\,{\sc ii}] emission; \citealt{greenawalt+98}) in the nucleus but little ionized gas in the outskirts. Deep IFU imaging of the optical lines (H$\alpha$ and [N\,{\sc ii}]) will be able to confirm this scenario.

\subsection{Molecular line ratios}
We assume that the HCN emission is emerging from the same region as the CN emission and then estimated the line ratio by converting the CARMA flux to the appropriate brightness temperature for the NRO\,45m.  We followed the same procedure for CS(2--1).  We find line ratios of  HCN(1--0)/CS(2--1)=1.2, CN(1--0)/HCN(1--0)=2.1 and CN(1--0)/CS(2--1)=3.3. Comparing different transitions of different species is not  ideal because one is tracing a mixture of excitation and chemistry.  In spite of this, it seems safe to say that CN is more abundant than HCN and CS in the central kpc of NGC\,5195.  

Using RADEX \citep{radex}\footnote{\href{http://home.strw.leidenuniv.nl/~moldata/radex.html}{http://home.strw.leidenuniv.nl/$\sim$moldata/radex.html}}, and assuming a gas kinetic temperature of 65\,K (from the dust temperature derived by \citealt{smith82}), a HCN/CS ratio of about unity can be obtained with volume densities of about 10$^4$cm$^{-3}$ and assuming similar HCN and CS abundances. This may point to a very low fraction of dense gas (i.e.  gas with number density $n$\,$\gtrsim$\,10$^5$cm$^{-3}$) that is in agreement with the emission of CN and CS, as well as with the high ratio CO(1-0)/HCN(1-0) of 85.3.\footnote{A HCN(1--0)/CS(2--1) line ratio close to unity may also emerge from an even denser gas component with $n$\,$>$\,10$^4$\,cm$^{-3}$, but this would require that the CS abundances exceed those of HCN. We assume collisional excitation only, no significant optical depth effects and that the hyperfine structure lines of HCN(1--0) can simply be added (see discussion in \citealt{aalto+15}).}

Assuming that CN is tracing PDRs in NGC\,5195,  the UV fields created by the low star formation in the galaxy (see \S\ref{sec:sf}) could dissociate and fragment the molecular clouds, creating larger envelopes in which most of the CN would form. The CN(1--0) line is brighter than that of HCN. The radiative transfer of CN is not straight-forward, but if we assume collisional excitation the line ratio indicates higher CN abundances that that of HCN.  Since CN can be a photodissociation product of HCN this may be the result of UV irradiation of fragmented dense clumps. However, one would then expect also to see a strong HCO$^+$ line. There are some star forming regions with X(CN)\,$>$\,X(HCN)\,$>$\,X(HCO$^+$) in (for example) the Orion ridge \citep{blake+86} but instead this ratio may emerge from the region around an AGN.  \citet{lepp+96} suggested that X(CN)\,$>$\,X(HCN) in X-ray irradiated regions. It is possible that (at least a significant fraction of) the high density tracer emission is emerging from the nucleus instead of being scattered in a post-starburst region.


Apart from the species mentioned before,  we  marginally detect CH$_3$CN$(6_k-5_k)$ (located between 200 and 350\,km~s$^{-1}$ at the red side of $^{13}$CO in Fig.~\ref{fig:spectra}). Its integrated flux, of 1.3 Jy~km~s$^{-1}$ is, however, below a 3$\sigma$ detection. This molecule is also a tracer of dense gas ($n_{\rm crit}$[CH$_3$CN$(6_k-5_k)]\sim10^5$cm$^{-3}$), similar to those of CN(1--0) and CS(2--1). However, CH$_3$CN is easily dissociated by UV fields \citep{aladro+13}, and thus it typically resides in the  cores of the molecular clouds. Based on that, the ratio between CN and CH$_3$CN would be an rough indicator of the amount of dense gas residing in the envelopes versus the inner parts. We obtain a tentative ratio of CN/CH$_3$CN=9.0. Interestingly the circumnuclear regions of the Seyfert galaxies NGC\,1097 and NGC\,1068 \citep{martin+15,aladro+13} show luminous CH$_3$CN emission. High resolution observations will reveal if also NGC\,5195 has a circumnuclear region of dense gas irradiated by an AGN. Such a structure may have a complex chemistry with suppressed HCO$^+$ while CN abundances can be high and also those of CH$_3$CN in warm, shielded regions.

\section{Past and present star formation in NGC\,5195}
\label{sec:sf}

\subsection{The stellar populations of NGC\,5195} 
\label{sec:stelpop}
In order to classify the stellar population of NGC\,5195, we used the Penalized Pixel Fitting software ({\sc ppxf}; \citealt{ppxf})\footnote{\href{http://www-astro.physics.ox.ac.uk/~mxc/software/\#ppxf}{http://www-astro.physics.ox.ac.uk/$\sim$mxc/software/{\#}ppxf}} applied to the nuclear spectrum\footnote{\href{http://data.sdss3.org/spectrumDetail?mjd=53063&fiber=527&plateid=1463}{http://data.sdss3.org/spectrumDetail?mjd=53063\&fiber=527\&plateid=1463}} of NGC\,5195 from the SDSS DR12 \citep{sdssdr12}. The 3$''$ spectral fiber is ideal for this study, as spectra of NGC\,5195 that are more extended than the nuclear region tend to be contaminated by the spiral arm of M51a (see: Figure~13.64 of \citealt{moustakas+10}). Integral field spectroscopy would be able to differentiate between NGC\,5195 and M51a, to gain a clearer picture of the integrated stellar population properties of NGC\,5195, but is beyond the scope of this paper.

We used the stellar templates from the {\sc miles} library \citep{miles}, normalized to a uniform \citep{salpeter55} initial mass function (IMF), using models representing 50 stellar ages from 0.06--17\,Gyr, and 4 metallicities (-0.71\,$<$\,{\em Z}\,$<$0.22), for a total of 200 models. We included regularization within {\sc ppxf} ({\tt regul}\,=\,0.004) to allow us to understand the range of possible stellar populations that would match the nuclear spectrum. Figure~\ref{fig:stelpop} shows the probability distribution function of the {\sc ppxf} model fitting, summed over the metallicity axis. The nucleus of NGC\,5195 clearly shows two stellar populations: 80\% of nuclear stars are old ($\gtrsim$\,10\,Gyr) and 20\% are of intermediate age ($\approx$\,1\,Gyr), consistent with A-stars. 

As Figure~\ref{fig:stelpop} shows, there is some uncertainty to the age of the intermediate-aged stellar population, mostly based on degeneracies of metallicity, but also likely because the SDSS fiber only subtends a small portion (144\,pc) of the nucleus. This does put a significant constraint on the interaction that took place between NGC\,5195 and M51a, suggesting when enhanced star formation was present in NGC\,5195. Given that star formation is known to increase even at large radii during gravitational encounters \citep{scudder+12,moreno+15}, it is likely that this stellar population represents the enhanced star formation that was taking place throughout the encounter (which was suggested to take place $\approx$\,1/2\,Gyr ago; \citealt{salo+00,dobbs+10}). The spectral stellar population fit also agrees within errors with the photometrically-derived stellar populations \citep{mentuch+12}, and might provide another constraint to model the complex history of interaction that has occurred in the M51 system.

\subsection{Integrated star formation}
\label{sec:int_SF}
The star formation rate of NGC\,5195 was calculated by \citet{lanz+13} using {\sc magphys} \citep{magphys} to model the far-UV to far-IR SED, measuring it to be 0.142~$M_\odot$\,yr$^{-1}$.  To estimate the contribution of the M51a to the \citet{lanz+13} SFR, we defined a region based on the 8\micron\ PAH emission from the foreground arm and measured the 70\micron\ and 8\micron\ photometry excluding the region. We find that 8\% of the PACS\,70\micron\ flux and 18\% of the IRAC\,8\micron\ flux are contained in the excluded region. While this region may also contain some emission from NGC\,5195, the strong correlation of the luminosities in these bands to SFR enable us to estimate that 10--20\% of the \citet{lanz+13} SFR may be due to M51a, revising the integrated SFR down to 0.116\,M$_\odot$\,yr$^{-1}$.

Using the revised SFR, and dividing by the moment0 area from Table~\ref{tab:mol_properties}, we derive a star formation rate surface density $\Sigma_{\rm SFR}$\,$\approx$\,0.105~$M_\odot$\,yr$^{-1}$\,kpc$^{-2}$ (renormalized to our chosen distance to NGC\,5195 and use of a Salpeter initial mass function; \citealt{salpeter55}).

To determine if the molecular gas is inefficient globally, we investigate where NGC\,5195 fits on the Schmidt-Kennicutt (S-K) star formation surface density - gas surface density relation \citep{schmidt59,ken98,kennicutt+12}:
\begin{equation}
\frac{\Sigma_{\rm SFR}}{M_\odot\,{\rm yr^{-1}\,kpc^{-2}}} = 2.5\times10^{-4}\left(\frac{\Sigma_{\rm H_2}}{M_\odot\,{\rm pc}^{-2}}\right)^{1.4}
\end{equation}
\noindent If we use the average molecular surface density calculated from our \twco\ observations, $\langle\Sigma_{\rm H_2}\rangle = 87$\,$\pm$\,$2$\,$M_\odot$\,pc$^{-2}$, we predict an average $\Sigma_{\rm SFR,predicted}$\,$\approx$\,0.13\,$M_\odot$\,yr$^{-1}$\,kpc$^{-2}$, in good agreement with the measured $\Sigma_{\rm SFR}$. We therefore conclude that the global molecular gas in NGC\,5195 is efficiently forming stars. Two caveats to this analysis are (1) that the \citet{lanz+13} star formation could have been contaminated by the spiral arm in M51a (seen in PAH emission in Figure~\ref{fig:pah+co}) and (2) that the intermediate-aged stellar population could be providing a non-zero contribution to the far-IR luminosity. We further investigate the star formation rate to determine if either of these caveats play a significant role.

\subsection{Radio-determined star formation}
\label{sec:radio_SF}
We also evaluated the SFR indicated by the centimeter- and millimeter-wave radio continuum emission using the following calibration
to the radio-SFR relation from \citet{murphy+11,murphy+13}:

\begin{equation}
\centering
\label{eqn:sfr_FF}
\begin{split}
\left(\frac{\rm SFR_\nu}{\rm M_\odot~yr^{-1}}\right) = 10^{-27} \\
		\left[2.18\left(\frac{T_{\rm ex}}{\rm 10^4\,K}\right)^{0.45}
		\left(\frac{\nu}{\rm GHz}\right)^{-0.1}+15.1\left(\frac{\nu}{\rm GHz}\right)^{\rm \alpha_{NT}}\right] \\
		\times\left(\frac{L_\nu}{\rm ergs\,s^{-1}\,Hz^{-1}}\right)
\end{split}
\end{equation}

\noindent with excitation temperature $T_{\rm ex}$\,=\,10$^4$\,K and the non-thermal spectral index $\alpha_{\rm NT}$\,=\,-0.8. From the sensitive interferometric imaging provided by the SINGS project \citep{sings,braun+07}, NGC\,5195 has an integrated 1.4\,GHz flux density of $\sim$\,17 mJy within the \twco-detected region.  This corresponds to a 1.4\,GHz SFR of $\sim$\,0.14\,$M_\odot$~yr$^{-1}$, slightly higher than the revised \citet{lanz+13} integrated SFR.  We therefore conclude that the centimeter-wave radio emission associated with NGC\,5195 is predominantly produced by SF, rather than radio-loud AGN activity.  This suggests that, if the nucleus of NGC\,5195 is indeed currently active (as the molecular lines seem to suggest), it may be operating in the so-called radio-quiet or ``high-excitation'' mode characteristic of efficiently accreting massive black holes \citep{heckman+14}, and consistent with the weak X-rays reported by \citet{schlegel+16}.  Alternatively, AGN signatures at optical and X-ray wavelengths previously reported by \citet{moustakas+10} and \citet{terashima+04} respectively may in fact be related to stellar processes or shocks occurring in the vicinity of the NGC\,5195 nucleus \citep{lisenfeld+10}.  If such a scenario in which the massive black hole in the center of NGC\,5195 is in a quiescent state is indeed true, it's not clear what mechanism is responsible for halting AGN fueling given the abundant availability of cold gas.  Additional detailed studies of the gas kinematics in the ambient environment of the NGC\,5195 nucleus will be needed to disentangle these possibilities in the future.

At the level of the integrated flux density of $\sim$\,1\,mJy of the compact 106\,GHz emission, the SFR predicted by Equation~\ref{eqn:sfr_FF} is $\sim$\,0.07\,$M_\odot$\,yr$^{-1}$. This is a lower limit, as it is likely that CARMA resolved out the diffuse, lower level emission (seen in the 1.4\,GHz map in \citealt{braun+07}). Even with this caveat, the predicted SFR based on the millimeter continuum emission is similarly consistent with a pure SF origin.

\subsection{Resolved star formation}
\label{sec:resolved_SF}
In order to correctly differentiate star formation taking place in the molecular gas in NGC\,5195, we use the reduced 70$\mu$m image (see: \S\ref{sec:ancillary}) from  {\em Herschel} PACS \citep{pacs}. The map was then converted to a $\Sigma_{\rm SFR}$ map using equation 21 from \citet{calzetti+10}. We were able to directly compare this map to the $\Sigma_{\rm H_2}$ map, which we derived from the \twco(1--0) using Equation~\ref{eqn:sigh2}, after converting the unclipped integrated intensity map using the Kelvin per Jansky factor for the CARMA map (listed in Table~\ref{tab:mol_properties}). We then re-gridded the 70$\mu$m map to match the \twco\ map using the {\sc idl} task {\tt hastrom}\footnote{\href{http://idlastro.gsfc.nasa.gov/ftp/pro/astrom/hastrom.pro}{http://idlastro.gsfc.nasa.gov/ftp/pro/astrom/hastrom.pro}}. We used the clipped moment maps (Figure~\ref{fig:moments}) to define the comparison region (which contains ``real'' emission far above the noise of the map, with a total area of 3.3\,kpc$^2$).

\begin{figure}[t]
\includegraphics[width=0.49\textwidth,clip,trim=0cm 0cm 0cm 0cm]{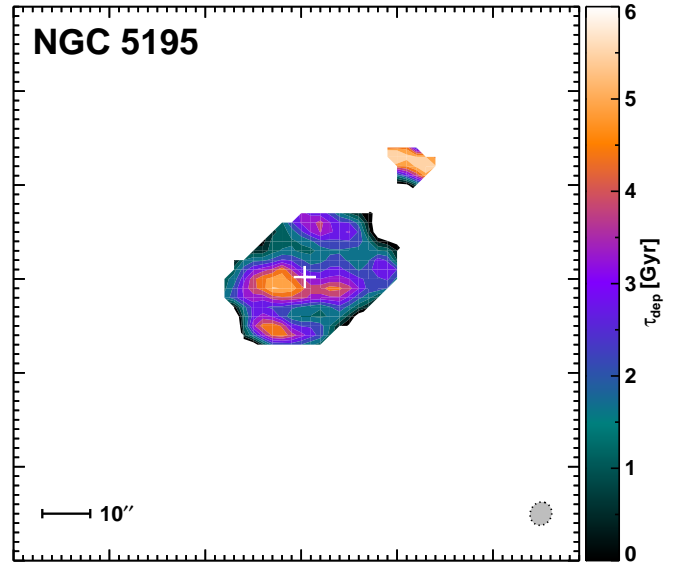} \vskip -1.5mm
\caption{The depletion time map for gas in NGC\,5195. The depletion time was calculated by dividing the matched $\Sigma_{\rm H_2}$ by $\Sigma_{\rm SFR}$. The CARMA beam is shown in the bottom right (gray). The mean depletion time for the gas in NGC\,5195 is $\langle\tau_{\rm dep}\rangle$ = 3.08\,Gyr. The position of the radio peak is shown as a white cross, which is not spatially coincident with the highest $\tau_{\rm dep}$, but is in its vicinity.}
\label{fig:tdep}
\end{figure}

We first calculated the total SFR by summing the $\Sigma_{\rm SFR}$ corresponding to \twco\ emission, SFR$_{\rm tot}$\,=\,0.072\,M$_\odot$\,yr$^{-1}$, that is a factor of 1.6 smaller than the revised SFR in \S\ref{sec:int_SF}, and consistent with the 106\,GHz free-free estimate for the SFR. This is possibly due to the smaller area subtended by the detected \twco\ emission. We then calculated the resolved depletion time for NGC\,5195 by dividing the matched $\Sigma_{\rm H_2}$ by $\Sigma_{\rm SFR}$, shown in Figure~\ref{fig:tdep}. The average depletion timescale (taken only from unmasked pixels with positive values) is 3.08\,Gyr (with a range between 1 and 6\,Gyr). 

When we include Helium, this depletion time increases to 4.31\,Gyr. This depletion timescale is a factor of $\approx$2 larger than what is found in normal star-forming galaxies \citep{bigiel+11,saintonge+11,leroy+13}. On the other hand, this depletion time is more consistent with the molecular gas depletion time found for ATLAS$^{\rm 3D}$ early-type galaxies \citep{davis+14}. Given that NGC\,5195 is an early-type galaxy, the depletion timescale consistent with early-type galaxies is fitting.

Overall, despite having taken part in the interactions with M51a a few hundred Myr ago, the gas in NGC\,5195 appears to have resettled and returned to forming stars at normal efficiency (for an early-type galaxy). This provides an update to the observations of \citet{kohno+02}, who suggested that the molecular disk was gravitationally stable against collapse, suppressing star formation in this system. Our data suggest that NGC\,5195 has returned to forming stars normally for its type (and its molecular gas content), which might include shear forces capable of reducing the star formation efficiency by $\sim$2 \citep{martig+13,davis+14}, consistent with other early-type galaxies. 

The level of star formation we currently observe in NGC\,5195 is consistent with a star formation history that includes an enhancement in star formation corresponding to a near approach and subsequent interaction with M51a (possibly including a super-efficient burst of star formation) that has declined for the past 1\,Gyr, followed by a resettling of the molecular disk and re-establishment of efficient, normal mode star formation consistent with its early-type classification.

\section{Conclusions}
\label{sec:conc}
We used CARMA to map four molecular gas tracers in NGC\,5195, detecting \twco(1--0), \thco(1--0), CN(1--0) and CS(2--1), and measured the 106\,GHz continuum emission in this galaxy, in addition to detections of HCN(1--0) and HCO$^+$(1--0) with the OSO 20m. We also fitted the SDSS DR12 nuclear spectrum of NGC\,5195 with stellar population models using a smoothed star formation history.

\begin{enumerate}
\item We find that our detections of \twco(1--0), \thco(1--0) and HCN(1--0) are consistent with previous observations and we provide the first measurements of CN(1--0), CS(2--1), HCO$^+$(1--0) and 106\,GHz continuum for NGC\,5195.

\item \rco\ has been updated for NGC\,5195 to be 11.4$\pm$0.5, which is much more consistent with a settled molecular distribution in a typical star-forming galaxy than a ULIRG, which tend to have large \rco\ (due to their disrupted molecular gas distributions).

\item The CN(1--0) emission found in the center of the galaxy appears to have kinematics that are different from the other molecular gas tracers studied. We suggest that the CN is tracing a diffuse component of the molecular gas that is found along the {\em x}2 orbits perpendicular to the stellar bar.

\item The molecular line ratios suggest that the dense gas ($n$\,$\gtrsim$\,10$^5$~cm$^{-3}$) fraction is low.

\item The stellar population fit to the nuclear spectrum of NGC\,5195 contains an 80\% mass fraction of old ($\gtrsim$\,10\,Gyr) stars and a 20\% mass fraction of intermediate age ($\approx$\,1\,Gyr) stars, consistent (within uncertainties) with an enhancement in star formation taking place starting during first approach and through the recent interaction with M51a.

\item The centimeter and millimeter continuum observations provide evidence that NGC\,5195 does not contain a buried AGN (or the AGN must be weak both in X-rays and very radio quiet). The star formation rates determined using the radio emission also supports the claim that the molecular gas in NGC\,5195 is forming stars at normal efficiency.

\item The resolved star formation relation taken from the {\em Herschel} 70$\mu$m maps of NGC\,5195 is consistent with the molecular gas forming stars at the efficiency observed for early-type galaxies. 
\end{enumerate}

\noindent Although NGC\,5195 has undergone a substantial interaction with M51a in the recent past ($\sim$\,1/2\,Gyr), it appears in the intervening time that its molecular gas has re-settled into a disk and re-established efficient star formation for its morphological type.

\section*{Acknowledgements}
The authors thank Dr Henrik Olofsson for carrying out the observations with the 20m telescope for them. KA thanks the anonymous referee for an insightful report that has markedly improved the manuscript, and M. Dapr\`a for {\sc gildas} advice. Support for KA is provided by NASA through Hubble Fellowship grant \hbox{\#HST-HF2-51352.001} awarded by the Space Telescope Science Institute, which is operated by the Association of Universities for Research in Astronomy, Inc., for NASA, under contract NAS5-26555. SA acknowledges support from the Swedish National Science Council grant 621-2011-4143. LL acknowledges support for this work provided by NASA through an award issued by JPL/Caltech. KN  acknowledges support from NASA  through the Spitzer Space Telescope.

Support for CARMA construction was derived from the Gordon and Betty Moore Foundation, the Kenneth T. and Eileen L. Norris Foundation, the James S. McDonnell Foundation, the Associates of the California Institute of Technology, the University of Chicago, the states of California, Illinois, and Maryland, and the National Science Foundation. Ongoing CARMA development and operations are supported by the National Science Foundation under a cooperative agreement, and by the CARMA partner universities.  The 20m telescope is operated by Onsala Space Observatory (OSO), the Swedish National Facility for Radio Astronomy.

The National Radio Astronomy Observatory is a facility of the National Science Foundation operated under cooperative agreement by Associated Universities, Inc. This research has made use of the NASA/IPAC Extragalactic Database (NED) which is operated by the Jet Propulsion Laboratory, California Institute of Technology, under contract with the National Aeronautics and Space Administration. This work is based [in part] on observations made with the {\em Spitzer} Space Telescope, which is operated by the Jet Propulsion Laboratory, California Institute of Technology under a contract with NASA. {\em Herschel} is an ESA space observatory with science instruments provided by European-led Principal Investigator consortia and with important participation from NASA. Funding for the Sloan Digital Sky Survey IV has been provided by the Alfred P. Sloan Foundation, the U.S. Department of Energy Office of Science, and the Participating Institutions. SDSS-IV acknowledges support and resources from the Center for High-Performance Computing at the University of Utah. The SDSS web site is www.sdss.org.

\noindent{\em Facilities:} \facility{CARMA}, \facility{{\em Herschel}}, \facility{OSO:20m}, \facility{SDSS}, \facility{{\it Spitzer}}

\bibliographystyle{aasjournal}
\bibliography{../../master}

\end{document}